\documentclass[useAMS,usenatbib]{mn2e}

\usepackage{umlaut}
\usepackage{color}
\usepackage{graphicx}
\usepackage{graphics}
\usepackage{rotating}
\usepackage{amsmath}        
\usepackage{amssymb}        

\newcommand{\apjs}{ApJS}

\newcommand{\aaps}{A\&AS}

\newcommand{\perbeam}{\,beam$^{-1}$}

\title[HSA imaging of V404 Cyg]{Zooming in on a sleeping giant: milliarcsecond HSA imaging of the black hole binary V404 Cyg in quiescence}

\author[J.C.A.~Miller-Jones et al.]
 {J.C.A.~Miller-Jones,$^{1,2}$\thanks{email: jmiller@nrao.edu} E.~Gallo,$^{3,4}$
 M.P.~Rupen,$^5$ A.J.~Mioduszewski,$^5$ \and W.~Brisken,$^1$
 R.P.~Fender,$^{6,7}$ P.G.~Jonker,$^{8,9,10}$ and T.J.~Maccarone,$^6$\\
$^1$NRAO Headquarters, 520 Edgemont Road, Charlottesville,
 VA, 22903, USA\\
$^2$Jansky Fellow, National Radio Astronomy Observatory\\
$^3$Physics Department, Broida Hall, University of California, Santa
 Barbara, CA, 93106, USA\\
$^4$Chandra Fellow\\
$^5$NRAO, Array Operations Center, 1003 Lopezville Road, Socorro, NM
 87801, U.S.A.\\
$^6$School of Physics and Astronomy, University of Southampton,
Highfield, Southampton, SO17 1BJ\\
$^7$Astronomical Institute `Anton Pannekoek', University of Amsterdam,
 Kruislaan  403, 1098 SJ, Amsterdam, The Netherlands\\
$^8$SRON, Netherlands Institute for Space Research, 3584 CA Utrecht,
 The Netherlands\\
$^9$Harvard-Smithsonian Center for Astrophysics, Cambridge, MA 02138,
 USA\\
$^{10}$Astronomical Institute, Utrecht University, 3508 TA, Utrecht,
 The Netherlands
}
\begin{document}

\date{Accepted xxxxxxx. Received xxxxxxx; in original form xxxxxxx}

\pagerange{\pageref{firstpage}--\pageref{lastpage}} \pubyear{2008}

\maketitle

\label{firstpage}

\begin{abstract}
Observations of the black hole X-ray binary V404 Cyg with the very
long baseline interferometer HSA (the High Sensitivity Array) have
detected the source at a frequency of 8.4\,GHz, providing a source
position accurate to 0.3\,mas relative to the calibrator source.  The
observations put an upper limit of 1.3\,mas on the source size
(5.2\,AU at 4\,kpc) and a lower limit of $7\times10^6$\,K on its
brightness temperature during the normal quiescent state, implying
that the radio emission must be non-thermal, most probably synchrotron
radiation, possibly from a jet.  The radio lightcurves show a short
flare, with a rise time of $\sim30$\,min, confirming that the source
remains active in the quiescent state.
\end{abstract}

\begin{keywords}
X-rays: binaries -- radio continuum: stars -- stars: individual
(V404 Cyg) -- ISM: jets and outflows -- astrometry
\end{keywords}

\section{Introduction}
\label{sec:intro}

Accreting black hole X-ray binary systems spend the majority of their
time in the so-called ``quiescent'' state, a low-luminosity state
(below a few $\times 10^{-6}$ of the Eddington luminosity) whose X-ray
emission is characterised by a hard ($1.5<\Gamma<2.1$) power-law
spectrum \citep{McC06}.  Despite such highly sub-Eddington systems
being so numerous, their intrinsic faintness poses serious
observational challenges to any attempt to understand the nature of
the accretion flow in these systems.

Accreting black holes have been studied in much more depth in the
``hard'' state, a regime with a similarly hard spectrum but occurring
at higher X-ray luminosities (up to a few per cent of the Eddington
luminosity).  In this state, they show flat radio spectra,
\citep[e.g][]{Fen01} interpreted as synchrotron emission arising from
partially self-absorbed, conical jets \citep{Bla79,Hje88}.  Such
jet-like structures have been directly imaged in two systems,
GRS\,1915+105 \citep{Dha00} and Cygnus X-1 \citep{Sti01}, at X-ray
luminosities of 5 and 0.3 per cent of the Eddington luminosity
respectively.  With only two relatively luminous examples of
self-absorbed jets, it has not been possible to study the dependence
of the jet size and geometry with system parameters such as orbital
period, nature of the mass donor, accretion rate, and black hole mass.

On moving from the hard to the quiescent state, the radio properties
of X-ray binaries become progressively less constrained, owing to the
limitations of current instrumentation.  In particular, the questions
of whether radio-emitting outflows persist into quiescence, and if so,
what form they take, remain to be answered.  Quiescence is currently
defined \citep{McC06} only by the luminosity criterion $L_{\rm
x}<10^{33.5}$\,erg\,s$^{-1}$ (i.e.\ $L_{\rm x}/L_{\rm
Edd}<2.3\times10^{-6}(M_{\rm BH}/10M_{\odot})$, where $L_{\rm Edd}$ is
the Eddington luminosity and $M_{\rm BH}$ is the black hole mass),
although there are indications that the X-ray spectrum in quiescence
softens from the canonical power-law index $\Gamma\sim1.5$ of the hard
state to a value closer to 2 \citep{Gal06,Cor06,Bra07}.  However, at
such low luminosities, instrumental sensitivity begins to hamper
efforts to characterise the behaviour of quiescent systems, and the
detailed spectral properties of the quiescent state remain to be
determined.  Extrapolation of the observed hard state behaviour
suggests that it is a reasonable assumption that jets should also
exist in the quiescent state.  This hypothesis is supported by the
detection of A0620-00 in the quiescent state in both the radio and
X-ray bands \citep{Gal06}, which demonstrated that the radio/X-ray
correlation seen in the hard state \citep{Gal03} remains unbroken down
to $10^{-8.5}L_{\rm Edd}$.  Further support is provided by the
observed flat radio spectrum in V404 Cyg \citep{Gal05}. However, the
evidence is circumstantial, and direct empirical proof for the
existence of such quiescent jets is still lacking.

\subsection{V404 Cyg}
V404 Cyg (GS\,2023+338) is the most luminous of the quiescent black hole
X-ray binaries, with a mean X-ray luminosity of $\sim 10^{33} (d/4{\rm
kpc})^2$\,erg\,s$^{-1}$ \citep{Gar01,Kon02,Hyn04,Bra07}.  Its high,
accurately-determined mass function, $6.08\pm0.06M_{\odot}$ \citep[the
most statistically significant measurement of a mass function above
$3M_{\odot}$;][]{Cas94}, implies that the compact object must be a
black hole.  The best-fitting system parameters indicate that the mass
donor is a K0 IV star \citep{Cas93} in a 6.5-d orbit with a black hole
of mass $12\pm2M_{\odot}$ \citep{Sha94}, although contamination of the
infrared spectrum by light from the accretion disc could reduce this
to $10M_{\odot}$, with a firm lower limit of $7.4M_{\odot}$
\citep{Sha96}.  \citet{Han92} measured a kinematic H{\sc i} distance
of $d\geq3.2$\,kpc.  Assuming Eddington-limited X-ray emission during
the 1989 outburst gives a maximum distance of $\leq3.7$\,kpc, whereas
the optical magnitude and reddening information suggests
$d=4.0^{+2.0}_{-1.2}$\,kpc \citep{Jon04}.  We will adopt the latter
value for the rest of this work.

The radio spectrum of V404 Cyg was measured to be flat, with a
spectral index of $\alpha=0.09\pm0.19$ (Gallo, Fender \& Hynes 2005),
defining $S_{\nu}\propto\nu^{\alpha}$.  Its flux density is $\sim
0.3$\,mJy at radio wavelengths (Gallo, Fender \& Pooley 2003),
although it is known to vary up to levels of 1.5\,mJy \citep{Hje00}.
The radio emission is believed to be of synchrotron origin
\citep{Gal05}, and the quiescent radio and X-ray luminosities are
consistent with an extrapolation of the non-linear radio/X-ray
correlation found for hard state X-ray binaries \citep{Gal03},
suggestive of jet emission.  However, the collimated nature of the
outflow remains to be proven.

\citet{Mun06} detected 4.5- and 8-\,$\mu$m infrared emission from V404
Cyg in excess of that expected from the Rayleigh-Jeans tail of the
companion star, which they attributed to emission from the accretion
disc.  However, \citet{Gal07} demonstrated that the broadband radio
through X-ray spectral energy distribution (SED) could be well fitted
with a maximally jet-dominated model, suggesting that the excess
infrared emission was consistent with being synchrotron emission from
a partially self-absorbed jet.

As the most luminous quiescent black hole X-ray binary, and with so much
circumstantial evidence for jets, V404 Cyg is the ideal candidate in
which to try to demonstrate empirically the existence of jets in a
quiescent accreting black hole system.  Since the source is so faint
($\sim0.3$\,mJy; other quiescent systems are fainter still), we
proposed for Very Long Baseline Interferometry (VLBI) observations
with the High Sensitivity Array (HSA) in an attempt to resolve the
jets, and as a pilot to check the feasibility of a future parallax
measurement to determine the source distance.

\section{Observations and data reduction}
\label{sec:obs}

V404 Cyg was observed for 4.25\,h with the HSA on 2007 December 2,
from 18:00:00 until 22:15:00 UT, under program code BG\,168.  We used
nine antennas of the Very Long Baseline Array (VLBA; the tenth
antenna, St.\ Croix, was down for maintenance) plus the Green Bank
Telescope (GBT), Effelsberg and the phased Very Large Array (VLA).
Eleven of the 24 available VLA stations were retrofitted EVLA
antennas.  We observed at 8.4\,GHz, in dual-circular polarization,
with a bandwidth of 32\,MHz per polarization.  Samples were quantized
to 2-bits, giving a total bit rate of 256\,Mb\,s$^{-1}$.  We used
J\,2025+3343, a 2-Jy source located only 16.6\,arcmin from the target,
as both the fringe finder and phase reference source.  We observed in
3-min cycles, spending 2\,min on the target and 1\,min on the
calibrator in each cycle.  Every 24\,min, we also made observations of
a check source, J\,2023+3153, the next nearest bright calibrator
(1.9\degr\ from the phase reference source), followed by a rephasing
of the VLA.

Data were reduced according to standard procedures within the National
Radio Astronomy Observatory's Astronomical Image Processing System
(AIPS) software \citep{Gre03}.  During the self-calibration of the
phase reference source, the initial model of the calibrator was
created using only the VLBA antennas, since the a priori calibration
(using system temperatures) for the VLBA antennas is better than that
for the larger dishes.  Amplitude self-calibration was then used to
correct the gains of the large dishes according to the VLBA-only
model.  Since the sensitivites on the typically long baselines
connecting the large dishes dominated the self-calibration solutions,
their absolute flux scale could not be sufficiently well determined
using standard calibration.  The visibilities on baselines from the
Los Alamos (LA) station to the VLA were compared to those at similar
({\it u,v}) co-ordinates between the LA and Pie Town (PT) stations,
and a global scaling factor of 0.89 was applied to correct the VLA
gains and bring the two sets of visibilities into agreement.  A
similar procedure was applied to baselines from the North Liberty (NL)
station to both the Hancock (HN) antenna and the GBT, although it was
found that no scaling was necessary for the GBT gains.  Finally, the
calibration derived on the phase reference source was transferred to
the target.  Owing to the low flux density of the target, no
self-calibration was performed.

\begin{figure}
\begin{center}
\includegraphics[height=\columnwidth,clip=]{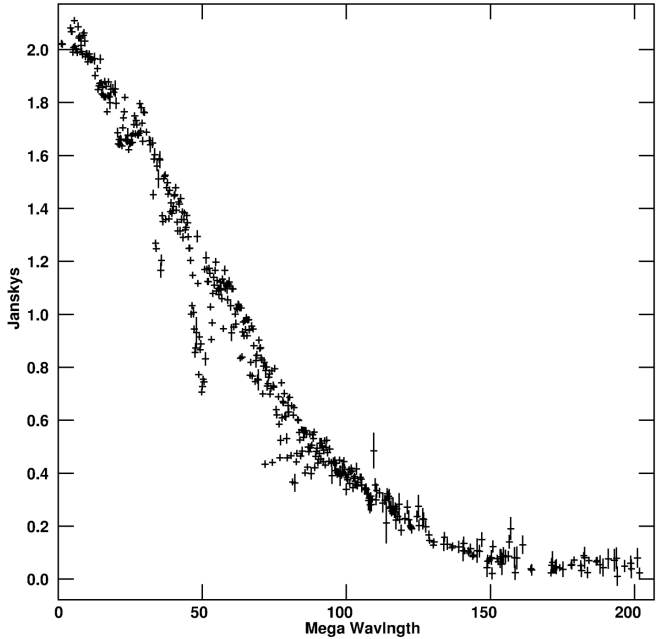}
\caption{Amplitude of the real part of the visibility plotted against
  radial distance from the centre of the {\it uv}-plane, for the phase
  reference source J\,2025+3343.  The shape of the plot is consistent
  with anisotropic interstellar scattering \citep[as parameterised by,
  e.g.,][]{Des01} giving rise to an elliptical scattering disc of
  size $2.2\times1.4$\,mas in PA 2\degr\ (see Section
  \ref{sec:scattering}), with the power-law index of the electron
  density fluctuations in the interstellar medium being close to 4, as
  found for J\,2025+3343 by \citet{Fey89}.
\label{fig:uvplt}}
\end{center}
\end{figure}

The extreme scattering towards the phase reference source
\citep{Fey89} meant that the achievable resolution was set by the
longest unscattered baseline to the calibrator, rather than the
expected instrumental resolution.  For this reason, the data from
Effelsberg had to be rejected.  Visibilities were scattered to zero
amplitudes on baselines longer than 150\,M$\lambda$
(Fig.~\ref{fig:uvplt}), whereas baselines to Effelsberg were between
120 and 290\,M$\lambda$.  Those in the range 120--150\,M$\lambda$ were
taken at the lowest elevations, and it was not deemed possible to
achieve accurate gain calibration for any of the data from this
station.

The VLA data were independently reduced and imaged, using standard
procedures within AIPS.  Weather conditions at the VLA were excellent
throughout the observations.  Amplitude calibration was performed
using observations of 3C\,48, adopting the flux density scale set by
the coefficients derived at the VLA in 1999.2, as implemented in the
31DEC08 version of AIPS.  While no proper polarization calibration
could be performed, the low polarization leakage of the VLA (of order
a few per cent) meant that we could obtain approximate constraints on
the linear ($P$) and circular ($V$) polarization of the source by
imaging the two internal IF pairs separately in all four Stokes
parameters $I$, $Q$, $U$ and $V$, and forming $P=(Q^2+U^2)^{1/2}$.

\section{Results}
\begin{figure}
\begin{center}
\includegraphics[height=\columnwidth,clip=]{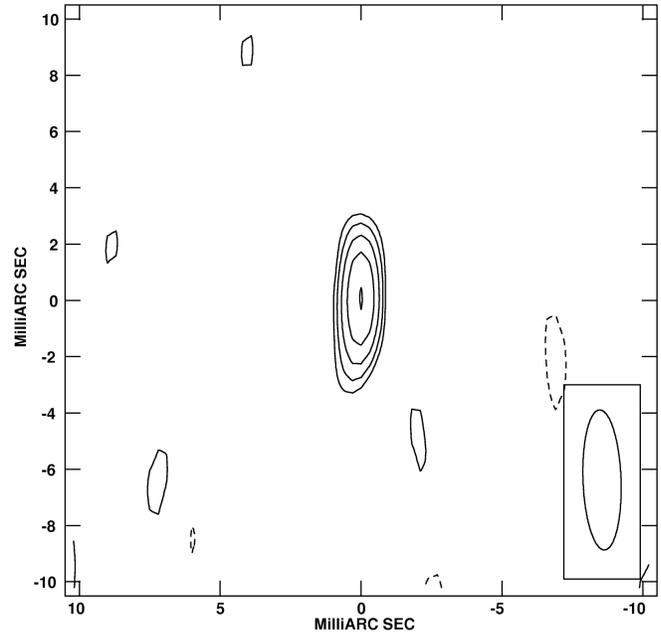}
\caption{Image of V404 Cyg.  The beamsize ($5.0\times1.4$\,mas$^2$ in
  PA 2\fdg0) is shown in the lower right.  Contours are at levels of
  $\pm(\sqrt{2})^n$ times the rms noise in the image
  (0.027\,mJy\,bm$^{-1}$), where $n=3,4,5,...$, with the lowest
  contour being at 0.078\,mJy\,bm$^{-1}$, and the peak flux density
  in the image being 0.322\,mJy\,bm$^{-1}$.  The source is
  unresolved.
\label{fig:v404_image}}
\end{center}
\end{figure}

V404 Cyg was detected by the HSA at a $12\sigma$ level with a mean
flux density of $0.32\pm0.03$\,mJy\,bm$^{-1}$ (here and in all the
following sections, unless specified, we quote $1\sigma$ error bars).
The source was unresolved down to a beamsize of
$5.0\times1.4$\,mas$^2$ in PA 2\fdg0 (Fig.~\ref{fig:v404_image}).  The
elongation is caused by the high sensitivities of the GBT and phased
VLA, with baselines to (and particularly between) those antennas
dominating the naturally-weighted sampling function.

We independently reduced the VLA data, to measure the total flux
density of the source and ascertain whether any emission was being
resolved out by the HSA.  The flux density measured by the VLA,
averaged over the length of the observation, was $0.47\pm0.03$\,mJy,
showing that the HSA is detecting most of the flux density, and
constraining the majority of the emission to come from a region
smaller than 5\,mas (Section \ref{sec:jet_size} discusses the size
constraint in more detail).

\subsection{Variability}
\begin{figure}
\begin{center}
\includegraphics[width=\columnwidth,angle=0,clip=]{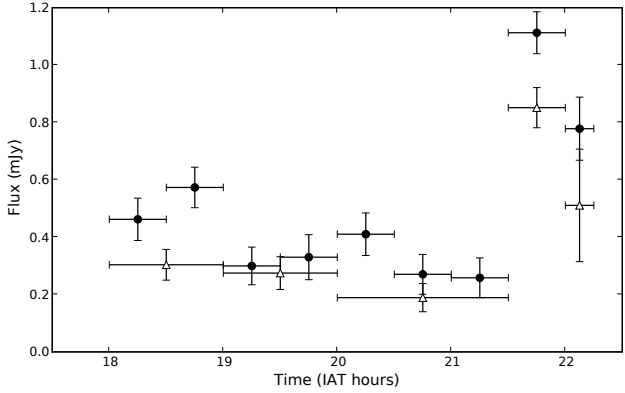}
\caption{VLA (filled circles) and HSA (open triangles) lightcurves
  during the HSA observations of V404 Cyg.  Note the flare at the end
  of the observations.
\label{fig:flare}}
\end{center}
\end{figure}

To search for shorter-timescale variability, the VLA data were split
into short (30-min) time segments and imaged.  The source was fit in
the image plane with a point source model during each 30-min interval,
and the resulting lightcurve is shown in Fig.~\ref{fig:flare}.

The measured HSA flux density is consistent with recovering $\sim70$
per cent of the total flux density measured with the VLA at all times.
When imaging the data from the flare at the end of the observations,
editing out the data from the phased VLA resulted in recovering a much
higher fraction of the total source flux ($1.00\pm0.14$\,mJy out of
$1.14\pm0.07$\,mJy), suggesting possible amplitude calibration or
phasing problems with the VLA.  However, a similar analysis performed
on the non-flare data showed no significant discrepancy between the
results with and without including the phased VLA.  Thus any problems
with the phased VLA were confined to the time of the flaring
observations.

\subsubsection{Flaring emission}
The source flux density as measured by the VLA rose by a factor of 3
during the flare at the end of the observing run, and there is also
the suggestion of a smaller flare earlier in the observation, at
18:45.  The rise time of the main flare was $<30$\,min, and it showed
no detectable circular or linear polarization (to a $5\sigma$ limit of
32 per cent), ruling out coherent stellar bursts from the donor as the
origin of the emission.  However, incoherent gyrosynchrotron stellar
flares with timescales of tens of minutes and much lower degrees of
circular polarization are common in RS CVn and Algol-type systems
\citep{Gud02}, and the peak radio luminosities of such flares can
reach levels of $10^{17}$--$10^{18}$\,erg\,s$^{-1}$\,Hz$^{-1}$
\citep[e.g.][]{Mut87,Ben94,Gar03}.  One such system, UX Ari, with a K0
subgiant in a 6.44-d orbital period with a G5 main sequence star
\citep{Str93}, makes for a good comparison to V404 Cyg in terms of the
rotation rate and stellar type of the flaring subgiant star.  The
maximum flare strength seen in this system in 3 years of radio
monitoring was 0.82\,Jy at 8.3\,GHz \citep{Ric03}, implying a peak
luminosity of $2.5\times10^{18}$\,erg\,s$^{-1}$\,Hz$^{-1}$, among the
highest observed to date in such systems.  The peak luminosity of the
flare seen in V404 Cyg was an order of magnitude higher still, making it
unlikely (if not impossible) that the observed flare in our data
originates from the donor star.

Splitting the HSA observations into 1-h chunks for imaging
(Fig.~\ref{fig:flare}), we found that the flare was also seen in the
high-resolution data.  Imaging only the data from during the flare
also showed that the source was unresolved, being smaller than the
beamsize of $4.7\times1.1$\,mas$^2$ in PA $-5$\fdg9.

\subsubsection{Non-flaring emission}
Taking only the data prior to the flare (18:00--21:30 UT), the HSA
measured a mean quiescent flux density of $0.24\pm0.03$\,mJy\perbeam,
and the source was unresolved with a beamsize of
$5.2\times1.4$\,mas$^2$ in PA 5\fdg5.  The VLA alone measured
$0.35\pm0.03$\,mJy\perbeam.  There was no evidence for any source
polarization with the VLA, down to a $5\sigma$ limit of 40 per cent
for both linear and circular polarization.

\subsection{Source position}

The shift in source position between quiescent and flaring states was
$0.13\pm0.31$\,mas ($1\sigma$ uncertainty), consistent with zero.
Using all the available data, the measured VLBI position, relative to
the phase reference source J\,2025+3343 (whose position was taken
to be\footnote{http://vlbi.gsfc.nasa.gov/solutions/2005f\_astro/2005f\_astro\_cat.txt} (J\,2000) 20$^{\rm h}$25$^{\rm m}$10\fs8421050(224)
33\degr43$^{\prime}$00\farcs21443(42)) was
\begin{equation}
\begin{split}
\notag
{\rm RA} &= 20^{\rm h}24^{\rm m}03\fs821269 \pm 0.000004\\
{\rm Dec.} &= +33\degr52^{\prime}01\farcs8988 \pm 0.0002\qquad{\rm (J\,2000)},
\end{split}
\end{equation}
where the quoted error bars are the purely statistical errors from
fitting in the image plane.  Imaging of the check source,
J\,2023+3153, phase referenced to J\,2025+3343, showed that the shift
in its position, relative to the catalogued position, was 0.34\,mas
in RA and 0.09\,mas in Dec., compared to uncertainties of 0.24 and
0.41\,mas respectively in the catalogued position of J\,2023+3153.
Since the check source was seven times further away from the phase
reference calibrator than the target source, we are therefore
confident that the systematic errors in the source position (e.g.\ due
to atmospheric phase gradients) are small and the positional
uncertainty is dominated by the signal-to-noise ratio on the target.

The unresolved nature of the source makes this system a good candidate
for a future parallactic distance determination, free from the extra
positional uncertainties introduced by source structure.  The current
measurement of the position would be used as an initial starting
point.  A distance of $4.0^{+2.0}_{-1.2}$\,kpc \citep{Jon04} would
imply a parallax of $\pi = 0.25^{+0.13}_{-0.08}$\,mas, which is
certainly feasible with current instrumentation.

\section{Jet size constraints}
\label{sec:jet_size}
\subsection{Image and visibility-based limits}
In order to put limits on the size of any extension, we fitted the
source with an elliptical Gaussian in both the image plane (with the
AIPS task JMFIT) and the {\it uv}-plane (using the AIPS task UVFIT and
the Caltech VLBI package Difmap).  Using the whole data set with all
available stations, the ($1\sigma$) upper limit on the FWHM of the
major axis was found to be $\sim 1.3$\,mas.  We note that a
sufficiently low-level extension would not show up in the fitting
results, so this is not a rigorous upper limit to the source size.

\subsection{Brightness temperature limits}
\label{sec:TBr}
Radio surface brightness may be parameterized in terms of the
temperature a blackbody would need to have in order to produce a
source of the observed size and flux density, at the observed
frequency, in the Rayleigh-Jeans limit.  This is the brightness
temperature, given by
\begin{equation}
T_{\rm b} = \frac{c^2 S_{\nu}}{2\nu^2 k_{\rm B} \Omega},
\end{equation}
where $c$ is the speed of light, $S_{\nu}$ the source flux density at
frequency $\nu$, $k_{\rm B}$ is Boltzmann's constant and $\Omega$ the
solid angle subtended by the source.  Our most stringent constraint
comes from the microflare, when the flux density measured by the HSA
was $1.0\pm0.1$\,mJy and fitting an elliptical Gaussian to the source
in the image plane gave an upper limit to the source size of
$1.8\times1.7$\,mas$^2$.  This gives a lower limit to the brightness
temperature of $T_{\rm b}>1.1\times10^7$\,K.  For comparison, for the
non-flaring data, fitting an elliptical Gaussian in the image plane
gave an upper limit to the size of $2\times0.6$\,mas$^2$ and a
consequent lower limit to the brightness temperature of
$7\times10^6$\,K.

Previous VLBI observations of Galactic X-ray binary jets have measured
brightness temperatures of $\geq10^9$\,K \citep[e.g][]{Dha00,Mir01}.
Should this be the case in V404 Cyg, we can use the flux density
during the flare to place an upper limit of $4.6\times10^{-19}$\,sr on
the solid angle subtended by the source, corresponding to a geometric
mean size of 0.16\,mas.  However, most resolved X-ray binary systems
have shown collimated jets, so as an example, for an axial ratio of
4:1 \citep[the definition of jets adopted by][]{Bri84}, the derived
solid angle would translate into a jet length of 0.3\,mas.

The inverse Compton catastrophe limits the brightness temperature of a
radio-emitting source to $<10^{12}$K, if it is not relativistically
boosted.  For the flux density of 1.0\,mJy\,\perbeam\ measured during
the flare, this limits the source size to $>5$\,$\mu$as
($>2$\,$\mu$as for the non-flaring emission with a flux density of
0.24\,mJy\perbeam).

\subsection{Variability limits}
The $<30$-min rise time of the flare corresponds to a distance of
$5.4\times10^{11}$\,m for a source moving at the speed of light, or an
angular size of 0.9\,mas if the motion is in the plane of the sky.
However, sources moving at small angles to the line of sight can show
apparent superluminal motion, with $\beta_{\rm app}>1$.  We have no
constraints on the speed or inclination angle to the line of sight of
the source, but constrain the size of the flaring region to be
$<0.9\beta_{\rm app}$\,mas.

Similarly, for the brightness temperature size constraints derived in
Section \ref{sec:TBr}, we can derive limits on the jet speed, for a
rise time of 30\,min.  For $T_{\rm B}\geq10^9$\,K, the jet speed is
limited to $<0.33c$, and for $T_{\rm B}<10^{12}$K, the jet speed is
$>830$\,km\,s$^{-1}$.  If the jet is to be moving close to the speed
of light, it would have to be more elongated, and have a lower
brightness temperature.

\subsection{Theoretical limits}
\citet{Hei06} derived an expression for the length of a flat-spectrum,
partially self-absorbed jet, as a function of basic jet parameters,
normalizing the relation using the VLBA measurements of the jet in
Cygnus X-1 \citep{Sti01}.  The observed angular extent of the jet,
$\mu_{50\%}$, within which 50 per cent of the flux density is
contained, can be written as
\begin{equation}
\mu_{50\%}\propto\left(\frac{S_{\nu}^{8}(\sin\theta)^{10}}{d\delta^2}\right)^{1/17}
\qquad {\rm mas},
\end{equation}
where $d$ is the source distance,
$\delta=[\Gamma(1-\beta\cos\theta)]^{-1}$ is the Doppler factor,
$\theta$ the inclination angle of the jet to the line of sight,
$\Gamma$ the Lorentz factor $(1-\beta^2)^{-1/2}$, and $\beta=v/c$ is
the jet velocity in units of the speed of light.  \citet{Sha94}
derived an inclination angle of $56\pm4$\degr\ for V404 Cyg, and we
measured a flux density of 0.3\,mJy.  \citet{Hei06} took values of
$\mu_{50\%}=3$\,mas, $S_{\nu}=12$\,mJy and $\theta=35$\degr\ for
Cygnus X-1.  With these parameters, a distance of 4\,kpc, and assuming
that the Doppler factor is similar in the two sources (since
$\mu_{50\%}$ scales as $\delta^{-2/17}$, the exact value is
unimportant) and that the disc is perpendicular to the jet, we predict
a jet length of 0.6\,mas in V404 Cyg.

\subsection{Previous constraints}
\citet{Mio08} also observed V404 Cyg with the VLBA and phased VLA.
From their non-detection at 15\,GHz, when the VLA lightcurve suggested
that an $8\sigma$ source should have been seen, they inferred that the
source was being resolved out, putting a lower limit on the source
size of 1\,mas.  Together with light-travel time arguments applied to
the flares they detected, they constrained the source size to lie in
the range 4--7\,AU.  While this is consistent with our upper limit of
1.3\,mas, it might initially appear to be difficult to reconcile with
the theoretical predictions of 0.3--0.6\,mas.  Either the models are
out by a factor of 2--3 (quite possible given some of the inherent
assumptions), or some unknown instrumental problem (possibly similar
to the apparent suppression of the flux density we found when using
the VLA during the flare) caused the source brightness to be reduced
below their expected $8\sigma$ detection.  In order to resolve the
jet, high-frequency, global VLBI would be required, to overcome the
scattering (which scales as $\nu^{-2}$) and achieve a resolution of
$\leq 1$\,mas.  Should the jets not be detected with higher-resolution
observations, it would present a serious challenge to the jet
interpretation of the radio emission from quiescent systems.

\section{Scattering}
\label{sec:scattering}
The line of sight to the Cygnus region is known to be highly
scatter-broadened \citep{Fey89}.  Fluctuations in the electron density
in the interstellar medium (ISM) modify the refractive index of the
plasma, inducing distortions in a wavefront travelling through the
medium.  This reduces the coherence of the electric field measured by
two separated antennas, reducing the amplitude of the visibilities
measured on long baselines, or, equivalently, giving rise to angular
broadening in the image plane.  \citet{Ric90} discusses other effects
of interstellar scattering, such as scintillation and temporal
broadening of pulsed emission.

\citet{Fey89} measured the scattering properties along the lines of
sight to several extragalactic calibrators in the Cygnus region,
including our phase reference source, J\,2025+3343 (i.e.\ B\,2023+336)
and our amplitude check source, J\,2023+3153 (i.e.\ B\,2021+317).  The
scattering is clearly non-uniform, and the scattering properties can
differ dramatically with small position offsets on the sky.  Between
our two calibrators, separated by only 1.9$^{\circ}$, the scattering
measure varies by a factor of five.  The measured scattering size of
J\,2025+3343 at 4.99\,GHz was $2.9\pm0.7$\,mas, scaling as
$\nu^{\gamma}$, with $\gamma=-1.97\pm0.29$.  At 8.4\,GHz, we measured
a scattering disc size of $2.2\times1.4$\,mas$^2$, fitting in either
the image or {\it uv}-plane.  This is significantly larger than the
previously-derived scattering size, although the size is known to
vary; \citet{Fey00} claimed evidence for a 50 per cent increase in
scattering disc size at all frequencies, in observations taken three
years after their original data.

\subsection{Source distance}
V404 Cyg lies behind the Cygnus superbubble.  \citet{Uya01} found that
the superbubble was not a single physical entity, but a projection of
unrelated features at different distances.  Nevertheless, the
structures are all believed to lie between 0.5 and 2.5\,kpc from us.
While the line of sight to V404 Cyg does not pass through any of the
known OB associations in the region, we can limit the location of the
major scattering screen to be within the same range of distances.
Assuming a single thin phase screen is responsible for the majority of
the scattering, then according to the relative distance of the
scattering screen and scattered source, we can predict the expected
scattering size,
\begin{equation}
\theta_{\rm gal} = (1-\frac{d_{\rm s}}{d})\theta_{\rm xgal},
\end{equation}
where $\theta_{\rm gal}$ is the scattering size of the galactic source
at distance $d$, $\theta_{\rm xgal}$ is the scattering size of the
background extragalactic source, and $d_{\rm s}$ is the distance from
the observer to the phase screen.  As demonstrated by \citet{Laz98},
the observed scattering size of a galactic source depends on the
relative distance of the source and the intervening phase scattering
screen.  The closer the source is to the scattering screen, the less
scattered the source will be.

The major axis of the calibrator scattering disc, 2.2\,mas, together
with the maximum major axis size of $\sim1.3$\,mas permitted by the
fits to the source data in the image and {\it uv} planes shows that if
the assumption of a thin phase screen is valid, and the scattering
measure is similar along the lines of sight to the source and
calibrator, then the lower limit to the source distance is
$d<2.4d_{\rm s}$.  This gives limits of $<1.2$ and $<6.1$\,kpc
respectively, for screen distances of 0.5 and 2.5\,kpc, consistent
with the range of 2.8--6.2\,kpc range derived by \citet{Jon04}.  While
this is not terribly constraining, it shows that lower-frequency
observations which resolved the scattering disc of V404 Cyg could
potentially provide an alternative, albeit model-dependent, estimate
of the source distance.

\subsection{Scintillation}
Given the high scattering towards the source, we should determine
whether the observed variability is intrinsic to the source, or
whether it could be caused by interstellar scintillation.  From the
scattering measure of $SM_{0} = 0.34\pm0.12$\,m$^{-20/3}$\,kpc
measured towards the phase reference source by \citet{Fey89}, and
assuming a scattering screen located at $d_{0} = 0.5$\,kpc, then the
standard scattering equations \citep[e.g.][]{Goo97} imply that we are
in the strong scattering regime (for which the cutoff frequency is
$\nu<100(d_{\rm s}/d_0)^{5/17}(SM/SM_0)^{6/17}$\,GHz with these
parameters).  Our observing bandwidth of 32\,MHz is too large to
detect diffractive scintillation (the decorrelation bandwidth is
$1.6(d_{\rm s}/d_0)^{-1}(SM/SM_0)^{-6/5}$\,MHz), and the source size,
limited to $>5$\,$\mu$as by the inverse Compton limit of $T_{\rm
b}=10^{12}$\,K, is also too large (diffractive scattering would
require $\theta<0.06(d_{\rm s}/d_0)^{-1}(SM/SM_0)^{-3/5}$\,$\mu$as).
Thus we are in the regime of refractive scintillation.  For intrinsic
source sizes between 5\,$\mu$as and 1.3\,mas (from the brightness
temperature limit and the measured upper limit on the source size from
the visibilities), the timescale of refractive scintillation is
between 140 and 1000\,h (scaling almost linearly with $d_{\rm s}$),
with a modulation index $0.02<m_{\rm R}<0.18$ (scaling as $\sim(d_{\rm
s}/d_0)^{-1/6}(SM/SM_0)^{1/2}$).  Thus the observed variability, an
increase by a factor of 3 on a timescale $<1$\,h, must be intrinsic to
the source.

\section{Flaring and the nature of the quiescent state}
Despite being in a ``quiescent'' state, V404 Cyg exhibits sub-orbital
variability in the optical \citep{Wag92} and infrared \citep{San96}
bands, with some evidence for a 6-h quasi-periodicity \citep{Cas93}.
\citet{Hyn02} found that variable photo-ionization of the accretion
disc by the X-ray source was responsible for the observed H$\alpha$
flares, a conclusion reinforced by observations of correlated X-ray
and H$\alpha$ variability \citep{Hyn04}.  The optical continuum flares
did not correlate as well with the X-ray flares, leading \citet{Hyn05}
to suggest that the continuum variability could be due to synchrotron
emission from a jet.  The source is known to vary in the radio band
\citep[e.g.][]{Hje00}, and \citet{Mio08} presented tentative evidence
for a similar 6-h radio periodicity.  The similar variability
characteristics (timescales and flare amplitudes) seen at the
different wavebands are suggestive of a common origin.

While certainly intrinsic to the source, and clearly a common
phenomenon in this system, the nature of the radio flaring is still to
be determined.  A thermal bremsstrahlung origin is ruled out, since it
would produce too high an X-ray luminosity for the system.  This
suggests that the radio emission is non-thermal (certainly plausible
given the relatively high brightness temperature), although a spectral
index measurement during a flare is needed to determine whether such
flares are optically thick or optically thin.  Due to a paucity of
simultaneous multiwavelength observations, any connection with the
optical continuum flares remains purely speculative.  We note,
however, evidence for a correlation between the optical and radio
bands is seen in the identical power-law decay timescale measured in
the two wavebands during the 1989 outburst \citep{Han92}.

The $<30$\,min rise time of the flare implies that the source of the
flaring emission must be of size $<4$\,AU, $\sim20$ times the size of
the binary orbit \citep{Sha94}.  \citet{Mio08} suggested three
possibilities for the origin of such flaring; collimated jets, shocks
in the circumstellar medium, or diffuse emission as seen in SS\,433
\citep{Blu01}.  The latter was on much larger scales (several hundred
AU), and was interpreted by the authors as bremsstrahlung emission,
for which reasons, we are inclined to favour the jet or shock
interpretations.

The flat radio spectrum seen in the quiescent state of V404 Cyg
\citep{Gal05} implies optically-thin thermal or optically-thick
synchrotron emission.  Thermal emission is ruled out due to
overpredicting the observed X-ray emission.  Optically-thick
synchrotron emission is typically interpreted as a partially
self-absorbed conical jet in X-ray binary systems, which would make
flaring in the jet, possibly due to internal shocks, a potential
candidate for the observed variability.  Other circumstantial evidence
for the existence of jets in the quiescent state is seen in the
correlation between the X-ray and radio luminosities of hard-state
black hole X-ray binaries at X-ray luminosities below $L_{\rm
x}\sim0.01L_{\rm Edd}$ \citep{Gal03}.  This correlation was later
demonstrated to hold across the entire black hole mass spectrum
(Merloni, Heinz \& di Matteo 2003; Falcke, K\"ording \& Markoff 2004),
and simultaneous X-ray and radio observations of A0620-00
\citep{Gal06} demonstrated that this correlation persists in
quiescence, down to X-ray luminosities as low as $10^{-8.5}L_{\rm
Edd}$.  The resolved jets seen in GRS\,1915+105 and Cygnus X-1 at the
high-luminosity end of the correlation then make it plausible that
jets are also present at much lower luminosities.  However, there are
several caveats to this argument.  A number of sources have recently
been found with lower than expected radio luminosities
\citep{Cor04,Cha06,Cad07,Rod07}, calling into question the
universality of the correlation \citep{Gal07a}.  Also, in the absence
of direct imaging, nothing is known about the extent and degree of
collimation of such quiescent jets, should they exist.  The inferred
larger inner disc radii derived in the quiescent state \citep{McC03}
could potentially affect the collimation of the outflow.  While the
SED of the quiescent source A0620-00 has been well-fitted with a
maximally jet-dominated model \citep{Gal07}, other models cannot be
ruled out.  Extended, flat-spectrum radio emission not arising from a
jet has been seen in the relatively uncollimated expanding structures
observed in CI Cam \citep{Mio04} and RS Oph \citep{Rup08} and the
equatorial emission of SS\,433 \citep{Blu01}.  Thus the true
morphology of the emission in V404 Cyg cannot be definitively
ascertained without high-resolution radio imaging.

\citet{Han92} suggested that the short-timescale variability in the
radio lightcurves of the 1989 outburst of V404 Cyg could be
interpreted as shocks at the ends of jets.  Circumstellar shocks have
been directly imaged in the high-mass X-ray binary CI Cam
\citep{Mio04}.  This system has a very dense stellar wind, which is
thought to have smothered the jets ejected during the outburst, giving
rise to an expanding shell of radio emission seen on VLBI scales.
While the wind in V404 Cyg is likely to be less strong, the size
constraint on the flares of V404 Cyg implies that they are occurring
closer to the binary than the shock seen in CI Cam, where any wind
from the donor star would be densest.  With the available
single-frequency, unresolved radio image, we cannot hope to pin down
the location of the shocks.  Ultimately, some explosive event is
required to cause the radio flares, and whether they are shocks within
or at the ends of jets cannot yet be determined.

As the most luminous black hole X-ray binary in quiescence, V404 Cyg is
the only source in which such flaring activity has been detected in
the quiescent state.  The question of whether the flares are unique to
this source, or are commonplace in such systems, has implications for
the nature of the accretion process at low luminosities.  Owing to the
limitations of current instrumentation, it has not been possible to
observe other quiescent systems with sufficient time resolution in the
radio band to detect such flares.  Models fitting the SEDs of
quiescent sources often assume a non-varying source, such that
non-simultaneous observations in different bands could be used to
build up the SED \citep[e.g.][]{Gal07}.  If flaring emission is
ubiquitous in these sources, the need for strict simultaneity will
have to be more rigorously enforced.

Observations of XTE J\,1118+480 and GX 339-4 in their hard states
(currently differentiated from the quiescent state only by a fairly
arbitrary luminosity criterion) have shown evidence for fast
variability in the optical and X-ray regimes \citep{Mot83,Hyn03}.  In
XTE J\,1118+480, both the autocorrelations and the cross-correlation
between the two wavebands were inconsistent with X-ray reprocessing
and more indicative of optical synchrotron variability \citep{Kan01},
which Malzac, Merloni \& Fabian (2004) were able to explain with a
jet-dominated model.  The timescales are shorter in this source than
seen in V404 Cyg however, and there were no high-time resolution radio
data for comparison (although we note that we would not expect to see
radio variability on timescales shorter than a few minutes, since any
variations occurring at the base of a jet would be smoothed out
further downstream where the source became optically thin in the radio
regime).  A better comparison is the quiescent black hole in the
centre of our Galaxy, Sgr A*, accreting at $\sim10^{-8}L_{\rm Edd}$.
It shows radio and infrared flaring activity (although any connection
between the two bands is as yet unclear), and the radio flaring has
been explained as adiabatic expansion of a self-absorbed transient
population of relativistic electrons \citep[e.g][]{Yus06}.

There is thus significant evidence that flaring activity in or at the
end of a jet is a feature of the quiescent state of accreting black
hole systems.  Whether there is a truly steady underlying jet as
assumed by standard jet models \citep{Bla79} or whether the emission
is composed of multiple overlapping flares, or some more uncollimated
structure, remains to be determined.  More sensitive instruments such
as the Expanded VLA (EVLA), e-MERLIN and the upgraded VLBA are
necessary to probe these short-timescale flares and determine the
nature of the quiescent emission.

\section{Conclusions}
We have detected the quiescent black hole X-ray binary system V404 Cyg
with the High Sensitivity Array at 8.4\,GHz.  The unresolved source
has a brightness temperature of $>10^7$\,K, and the emission is
inferred to be non-thermal synchrotron radiation.  Our detection
provides the most accurate source position to date, which can serve as
an initial point for future proper motion studies.  We have also
constrained the length of the jet in this system to be $<1.3$\,mas,
consistent with previous measurements and theoretical expectations.

A small flare was detected with both the VLA and the HSA, with a rise
timescale of 30\,min, in which the source flux density rose by a
factor of 3.  This flare is certainly intrinsic to the source, and
cannot be caused by interstellar scintillation.  This is further
evidence that source flux densities in the quiescent state are not
stable, mandating the use of strictly simultaneous observations
when constructing and modelling SEDs to determine the contribution of
different components (such as jets) to the observed emission in
different wavebands.

The detection of an unresolved quiescent black hole X-ray binary
system opens the way for further high-resolution observations.  If the
inferred jets indeed exist, and are flat-spectrum, as has been
proposed, then higher-frequency observations would be well-placed to
resolve the jets, if the theoretical predictions for the jet length
are correct.  Furthermore, the unresolved source at 8.4\,GHz is an
ideal candidate for a parallactic distance determination, free from
the uncertainties associated with the source structure seen when such
systems are observed with VLBI during outburst.

\section*{Acknowledgments}
This research has made use of NASA's Astrophysics Data System.
J.C.A.M.-J.  is a Jansky Fellow of the National Radio Astronomy
Observatory.  E.G. is supported through Chandra Postdoctoral
Fellowship grant number PF5-60037, issued by the Chandra X-Ray Center,
which is operated by the Smithsonian Astrophysical Observatory for
NASA under contract NAS8-03060.  The VLBA is a facility of the
National Radio Astronomy Observatory which is operated by Associated
Universities, Inc., under cooperative agreement with the National
Science Foundation.

\label{lastpage}
\bibliographystyle{mn2e}

\end{document}